\documentclass[12pt,aps,prd,showpacs,amsmath,amssymb]{revtex4}
\input epsf
\begin{document}
\title{Electromagnetic form factors of the $\Lambda$ and $\Sigma$ baryons in an alternative baryonic current approach}
\author{Yong-Lu Liu and Ming-Qiu Huang}
\affiliation{Department of Physics, National University of Defense Technology, Hunan 410073, China}
\date{\today}
\begin{abstract}
Light-cone sum rules are used to investigate the electromagnetic form factors of the $\Lambda$ and $\Sigma$ baryons by using the Ioffe type interpolating currents.
The sum rules are affected to some extent by the choice of the interpolating currents from a comparison. Numerical calculations show that the magnetic form factor
can be well fitted by the dipole formula for $\Sigma$ but not for $\Lambda$. The magnetic form factor of $\Lambda$ approaches zero with the momentum transfer faster
than the dipole formula estimation.
\end{abstract}
\pacs{14.20.-c, 11.25.Hf, 11.55.Hx, 21.10.-k, 13.40.-f} \maketitle

\section{Introduction}
\label{sec1} Electromagnetic (EM) form factors are important observable parameters to probe the inner structure of the baryon and eventually understand the strong
interaction \cite{Arrington,Perdrisat}. Recent developments in experimental instruments provide opportunities to give new results on the EM form factors of the
nucleon \cite{Bourgeois,Jlab} (for a review see Refs. \cite{Arrington,Perdrisat} and references therein). The experiments show that data of the electric form factor
from the polarization transfer technique decrease faster than the dipole law for squared momentum transfer $Q^2$ \cite{Jlab,Melo}. Theoretically, there were a lot
of studies on the EM form factors of the nucleon in the past years. The recent status of the development can be found in Ref. \cite{Lenz1} and references therein.
Nevertheless, investigations on the EM form factors of the $\Lambda$ and $\Sigma$ baryons are much less than that of the nucleon as there are no experimental data
available, so further theoretical study on this issue is needed in order to understand the internal structure of these baryons.

Among the various theoretical investigations on $\Lambda$ and $\Sigma$, chiral perturbation theory and the chiral quark/soliton model have been used to study the EM
form factors at low momentum transfer \cite{Kubis,Kim}; the authors of Refs. \cite{Cauteren} have studied the electric and magnetic form factors of these baryons in
the relativistic constituent quark model. We have given the distribution amplitudes of $\Lambda$ and $\Sigma$ in Ref. \cite{DAs}, which makes it possible to use the
light cone sum rule (LCSR) approach to study the EM form factors of these baryons \cite{lcsr1,lcsr2,lcsr3,Colangelo}. LCSR is a nonperturbative method which
includes both the traditional QCD sum rule \cite{SVZ} and the hard exclusive process theory \cite{exclusive,exclusive2}. It has been widely used to investigate the
hadronic dynamics related to strong interaction \cite{Lenz,Aliev,Wang,WYM}.

In our previous papers \cite{DAs,EMff}, we have studied the EM form factors of the baryons using the interpolating currents provided by Chernyak and Zhitnitsky (CZ
current) \cite{Chernyak} in the framework of LCSR. It has been known that the hadron interpolating field is one of the most important ingredients in nonperturbative
field-theoretic investigations. As three-body composite particles, baryons have several independent interpolating currents with the same quantum numbers.
Particularly, for baryons with spin $1/2$, there are two independent interpolating fields with no derivatives \cite{Chung,Ioffe}. Linear compositions of these two
interpolators have been investigated for both light and heavy baryons within QCD sum rules \cite{Lee,IIG,WDW}. The author of Ref. \cite{Leinweber} shows that the
mixing of the two kinds of currents may affect the results for the nucleon. In Refs. \cite{Lenz,Aliev} it has been shown that the choice of the currents affects the
EM form factors greatly in the case of the nucleon. Thus it is expected that different interpolating currents may give information on different aspects of the same
process. The aims of the present paper are to study the EM form factors of the $\Lambda$ and $\Sigma$ baryons at moderately large momentum transfer in LCSR with the
choice of the Ioffe type interpolating current \cite{Ioffe}, and to estimate the magnetic moments from the magnetic form factors obtained.

We proceed as follows. Section \ref{sec2} is devoted to derive the light cone sum rules of the EM form factors for $\Lambda$ and $\Sigma$. Section \ref{sec3} is the
numerical analysis part. Summary and conclusion are given at the end of this section.  Appendix \ref{appendix} presents the distribution amplitudes of the baryons.

\section{Light-cone QCD Sum Rules for EM form factors}\label{sec2}

The most general expressions of spin $1/2$ baryonic EM form factors are defined by the matrix element of the EM currents between baryon states:
\begin{equation}
\langle B(P',s')|j_\mu^{em}(0)|B(P,s)\rangle=\bar B(P',s')[\gamma_\mu F_1(Q^2)-i\frac{\sigma_{\mu\nu} q^\nu}{2M}F_2(Q^2)]B(P,s),\label{form}
\end{equation}
where $F_1(Q^2)$ and $F_2(Q^2)$ are Dirac and Pauli form factors, respectively. $B(P,s)$ denotes the baryon spinor with momentum $P$ and spin $s$, $M$ is the baryon
mass, $Q^2=-q^2=-(P-P')^2$ is the squared momentum transfer, and $j_\mu^{em}$ is the EM current relevant to the baryon. Experimentally speaking, the Dirac and Pauli
form factors can also be expressed by the magnetic and electric Sachs form factors $G_M(Q^2)$ and $G_E(Q^2)$:
\begin{eqnarray}
G_M(Q^2)&=&F_1(Q^2)+F_2(Q^2),\nonumber\\
G_E(Q^2)&=&F_1(Q^2)-\frac{Q^2}{4M^2}F_2(Q^2),
\end{eqnarray}
which describe the distributions of the magnetic current and the electric charge in the Breit frame. In particular, the normalization of these form factors at the
point $Q^2=0$ is given by the baryon magnetic moment and electric charge.

We have ever studied the EM form factors of the $\Lambda$ and $\Sigma$ baryons in Ref. \cite{EMff,DAs} using the following CZ current:
\begin{equation}
j_{CZ}(x)=\epsilon^{ijk}[q_1^i(x)C\!\not\! zq_2^j(x)]\gamma_5\!\not\! z s^j(x),
\end{equation}
in which $q_i(x)$ and $s(x)$ represent the quark fields, $C$ is the charge conjugation operator and $z$ is a light-cone vector which satisfies $z^2=0$. The coupling
of this current determines the normalization of the leading twist distribution amplitudes. However, as mentioned in the above section, the choice of the baryonic
interpolating current may affect the results of sum rules \cite{Lenz,IIG}. Furthermore, a usage of different interpolating currents in the LCSR application makes a
complement for the theoretical analysis. In this regard, we adopt the interpolating field similarly as that used for the nucleon by Ioffe \cite{Ioffe} to study the
EM form factors of $\Lambda$ and $\Sigma$. The coupling constant of the Ioffe type current determines the normalization of higher order twist distribution
amplitudes \cite{DAs,Braun}. In fact, only axial vector (vector) structures contribute to the sum rules for $\Lambda$ ($\Sigma$) with the usage of CZ current, while
the application of Ioffe type current may include contributions from both vector like and axial vector like distribution amplitudes. Hence it is of interest to
investigate the EM form factors by LCSR with the Ioffe type interpolating current, which is expected to approve our theoretical analysis and give new information
for the $\Lambda$ and $\Sigma$ baryons.
\subsection{LCSRs of EM form factors for $\Lambda$}

Following the standard philosophy in LCSR, the procedure begins with the correlation function:
\begin{equation}
T_\mu(P,q)=i \int d^4xe^{iq\cdot x}\langle
0|T\{j_\Lambda(0)j_\mu^{em}(x)\}|\Lambda(P)\rangle,\label{correlator1}
\end{equation}
where the interpolating current of the $\Lambda$ baryon field is chosen as Ioffe type one:
\begin{equation}
{j_\Lambda}(x)=\epsilon_{ijk}(u^i(x)C\gamma_5 \gamma_\mu d^j(x))\gamma_\mu s^k(x),
\end{equation}
and the coupling constant is defined by the matrix element of the interpolating current between the vacuum and the $\Lambda$ state:
\begin{equation}
\langle 0|j_\Lambda|\Lambda\rangle=\lambda_1M\Lambda(P),\label{norm}
\end{equation}
in which $M$ is the mass of the baryon. The EM current is
\begin{equation}
j_\mu^{em}(x)=e_u\bar u(x)\gamma_\mu u(x)+e_d \bar d(x)\gamma_\mu d(x)+e_s \bar s(x) \gamma_\mu s(x).
\end{equation}

The derivation of the sum rules needs to express the correlation function (\ref{correlator1}) both phenomenologically and theoretically. On the one hand, by
inserting a complete set of intermediate states with the same quantum numbers as those of $\Lambda$, using the definitions of Eq. (\ref{form}) and Eq. (\ref{norm}),
the hadronic representation of the correlation function is written as
\begin{eqnarray}
z^\mu T_\mu(P,q)&=&\frac{\lambda_1M}{M^2-P'^2}[2P\cdot zF_1(Q^2)+\frac{P\cdot z}{M}\!\not\! q_\perp
F_2(Q^2)\nonumber\\
&&+(F_1(Q^2)+F_2(Q^2))\!\not\! z\!\not\! q+\frac{q^2}{2M}\!\not\! zF_2(Q^2)]\Lambda(P)+...,\label{hadrep}
\end{eqnarray}
where $P'=P-q$, $\!\not\! q_\perp=\!\not\! q-\frac{pq}{pz} \!\not\! z$ and the dots stand for the higher resonance contributions. Here we contract the correlation
function with the light-cone vector $z^\mu$ to get rid of contributions proportional to $z^\mu$ which are subdominant on the light cone.

On the other hand, the correlation function can be expanded on the light cone at large Euclidean momenta $P'^2$ and $q^2=-Q^2$ in terms of the distribution
amplitudes which are presented in Appendix \ref{appendix}. As the usual procedure, the hadronic representation of the correlation function is matched with the QCD
calculations on the light-cone to obtain the sum rules. The correlation function (\ref{correlator1}) contains several Lorenz structures (as in Eq.(\ref{hadrep})),
and in principle all of them can provide information of the EM form factors. In the paper, terms proportional to $1$ and $\!\not\! q_\perp$ are chosen to obtain the
related form factors. After taking advantage of the dispersion relation and the hadron-quark duality approach, the light-cone QCD sum rules are given as the
following Borel transformed version:
\begin{eqnarray}
2\lambda_1F_1(Q^2)&=&e_u\Big\{\int_{\alpha_{10}}^1d\alpha_1\{-B_0(\alpha_1)-\frac{1}{\alpha_1}B_1(\alpha_1)
-\frac{M^2}{M_B^2}B_2(\alpha_1)+\frac{Q^2}{\alpha_1^2M_B^2}B_3(\alpha_1)\nonumber\\
&& +\frac{2M^2}{\alpha_1M_B^2}B_4(\alpha_1)\}e^{-\frac{s_1-M^2}{M_B^2}}+e^{-\frac{s_0-M^2}{M_B^2}}\frac{\alpha_{20}^2}{\alpha_{10}^2M^2+Q^2}\{-M^2B_2(\alpha_{10})
\nonumber\\
&&+\frac{Q^2}{\alpha_{10}}B_3(\alpha_{10})+\frac{2M^2}{\alpha_{10}}B_4(\alpha_{10})\}\Big\}\nonumber\\
&&+e_d\Big\{\int_{\alpha_{20}}^1d\alpha_2\{C_0(\alpha_2)+\frac{1}{\alpha_2}C_1(\alpha_2)+\frac{M^2}{M_B^2}C_2(\alpha_2)-\frac{Q^2}{\alpha_2^2M_B^2}C_3(\alpha_2)
\nonumber\\
&&+\frac{2M^2}{\alpha_2M_B^2}C_4(\alpha_1)\}e^{-\frac{s_2-M^2}{M_B^2}}+e^{-\frac{s_0-M^2}{M_B^2}}\frac{\alpha_{20}^2}{\alpha_{20}^2M^2+Q^2}\{M^2C_2(\alpha_{20})
\nonumber\\
&&-\frac{Q^2}{\alpha_{20}}C_3(\alpha_{20})+\frac{2M^2}{\alpha_{20}}C_4(\alpha_{20})\}\Big\}\nonumber\\
&&+2e_s\Big\{\{\int_{\alpha_{30}}^1d\alpha_3\{D_1(\alpha_3)+\frac{1}{\alpha_3}D_1(\alpha_2)
-\frac{m_sM}{\alpha_3M_B^2}D_2(\alpha_3)-\frac{Q^2}{\alpha_3^2M_B^2}D_1(\alpha_3)\nonumber\\
&&+\frac{M^2}{M_B^2}D_3(\alpha_3)+\frac{m_sM^3}{\alpha_3M_B^4}D_4(\alpha_3)\}e^{-\frac{s_2-M^2}{M_B^2}}+e^{-\frac{s_0-M^2}{M_B^2}}
\frac{1}{\alpha_{30}^2M^2+Q^2+m_s^2}\nonumber\\
&&\times\{-\alpha_{30}m_sMD_2(\alpha_{30})-Q^2D_1(\alpha_{30})+\alpha_{30}^2M^2D_3(\alpha_{30})+\frac{\alpha_{30}m_sM^3}{M_B^2}\nonumber\\
&&\times D_4(\alpha_{30})-\alpha_{30}^2M^2\frac{d}{d\alpha_{30}}D_4(\alpha_{30})\frac{\alpha_{30}m_sM}{\alpha_{30}^2M^2+Q^2+m_s^2}\}\Big\},
\end{eqnarray}
and the sum rule for $F_2(Q^2)$ is
\begin{eqnarray}
\lambda_1F_2(Q^2)&=&e_u\Big\{\int_{\alpha_{10}}^1d\alpha_1\frac{1}{\alpha_1}\{E_1(\alpha_1)+\frac{M^2}{M_B^2}E_2(\alpha_1)\}e^{-\frac{s_1-M^2}{M_B^2}}\nonumber\\
&&+e^{-\frac{s_0-M^2}{M_B^2}}E_2(\alpha_{10})\frac{\alpha_{10}M^2}{\alpha_{10}^2M^2+Q^2}\Big\}\nonumber\\
&&+e_d\Big\{\int_{\alpha_{20}}^1d\alpha_2\frac{1}{\alpha_2}\{F_1(\alpha_2)+\frac{M^2}{M_B^2}F_2(\alpha_2)\}e^{-\frac{s_2-M^2}{M_B^2}}\nonumber\\
&&+e^{-\frac{s_0-M^2}{M_B^2}}F_2(\alpha_{20})\frac{\alpha_{20}M^2}{\alpha_{20}^2M^2+Q^2}\Big\}\nonumber\\
&&+2e_s\Big\{\int_{\alpha_{30}}^1d\alpha_2\frac{1}{\alpha_3}\{G_1(\alpha_3)-\frac{M^2}{M_B^2}G_2(\alpha_3)+\frac{m_sM}{\alpha_3M_B^2}G_3(\alpha_3)\nonumber\\
&& -\frac{m_sM^3}{\alpha_3M_B^3}G_4(\alpha_3)\}
e^{-\frac{s_3-M^2}{M_B^2}}+e^{-\frac{s_0-M^2}{M_B^2}}\frac{\alpha_{30}}{\alpha_{30}^2M^2+Q^2+m_s^2}\nonumber\\
&&\times\{-M^2G_2(\alpha_{30})+\frac{m_sM}{\alpha_{30}}G_3(\alpha_{30})-\frac{m_sM^3}{\alpha_3M_B^2}G_4(\alpha_{30})\nonumber\\
&&+M^3\frac{d}{d\alpha_{30}}\frac{m_s}{\alpha_{30}^2M^2 +Q^2+m_s^2}G_4(\alpha_{30})\}\Big\},
\end{eqnarray}
where $M_B$ is the Borel parameter, $s_i=(1-\alpha_i)M^2+\frac{1-\alpha_i}{\alpha_i}Q^2+\frac{m_i^2}{\alpha_i}$, $m_{1,2}=0,m_3=m_s$, and $\alpha_{i_0}$ connects
with the continuum threshold $s_0$:
\begin{equation}
\alpha_{i_0}=\frac{-(s_0+Q^2-M^2)+\sqrt{(s_0+Q^2-M^2)^2+4(Q^2+m_i^2)M^2}}{2M^2}.\label{threshold}
\end{equation}
The following notations are used for convenience:
\begin{eqnarray}
B_0(\alpha_1)&=&\int_0^{1-\alpha_1}d\alpha_2(V_3-2A_1-3A_3)(\alpha_1,\alpha_2,1-\alpha_1-\alpha_2),\nonumber\\
B_1(\alpha_1)&=&(\widetilde V_1-\widetilde V_2-\widetilde V_3-\widetilde A_1+\widetilde A_2-\widetilde
A_3)(\alpha_1),\nonumber\\
B_2(\alpha_1)&=&(\widetilde V_3-\widetilde V_4-2\widetilde
A_1-3\widetilde A_3+\widetilde A_4+2\widetilde
A_5)(\alpha_1),\nonumber\\
B_3(\alpha_1)&=&(\widetilde V_1-\widetilde V_2-\widetilde
V_3+\widetilde A_1-\widetilde A_2+\widetilde
A_3)(\alpha_1),\nonumber\\
B_4(\alpha_1)&=&(\widetilde{\widetilde A}_1-\widetilde{\widetilde
A}_2+\widetilde{\widetilde A}_3+\widetilde{\widetilde
A}_4-\widetilde{\widetilde A}_5+\widetilde{\widetilde
A}_6)(\alpha_1),\nonumber\\
C_0(\alpha_2)&=&\int_0^{1-\alpha_2}d\alpha_1(V_3+2A_1+3A_3)(\alpha_1,\alpha_2,1-\alpha_1-\alpha_2),\nonumber\\
C_1(\alpha_2)&=&(\widetilde V_1-\widetilde V_2-\widetilde
V_3+\widetilde A_1-\widetilde A_2+\widetilde
A_3)(\alpha_2),\nonumber\\
C_2(\alpha_2)&=&(\widetilde V_3-\widetilde V_4+2\widetilde
A_1+3\widetilde A_3-\widetilde A_4-2\widetilde
A_5)(\alpha_2),\nonumber\\
C_3(\alpha_2)&=&(\widetilde V_1-\widetilde V_2-\widetilde
V_3-\widetilde A_1+\widetilde A_2-\widetilde
A_3)(\alpha_2),\nonumber\\
C_4(\alpha_2)&=&(\widetilde{\widetilde A}_1-\widetilde{\widetilde
A}_2+\widetilde{\widetilde A}_3+\widetilde{\widetilde
A}_4-\widetilde{\widetilde A}_5+\widetilde{\widetilde
A}_6)(\alpha_2),\nonumber\\
D_0(\alpha_3)&=&\int_0^{1-\alpha_3}d\alpha_1(\frac{m_s}{\alpha_3M}A_1+A_3)(\alpha_1,1-\alpha_1-\alpha_3,\alpha_3),\nonumber\\
D_1(\alpha_3)&=&(-\widetilde A_1+\widetilde A_2-\widetilde
A_3)(\alpha_3),\nonumber\\
D_2(\alpha_3)&=&(-2\widetilde A_1+\widetilde A_2-\widetilde
A_3-\widetilde A_4+\widetilde A_5)(\alpha_3),\nonumber\\
D_3(\alpha_3)&=&(\widetilde A_3-\widetilde
A_4)(\alpha_3),\nonumber\\
D_4(\alpha_3)&=&(\widetilde{\widetilde A}_1-\widetilde{\widetilde
A}_2+\widetilde{\widetilde A}_3+\widetilde{\widetilde
A}_4-\widetilde{\widetilde A}_5+\widetilde{\widetilde
A}_6)(\alpha_2),\nonumber\\
E_1(\alpha_1)&=&\int_0^{1-\alpha_1}d\alpha_2(V_1-A_1)(\alpha_1,\alpha_2,1-\alpha_1-\alpha_2),\nonumber\\
E_2(\alpha_1)&=&(\widetilde V_1-\widetilde V_2-\widetilde V_4-\widetilde A_1-\widetilde A_2-2\widetilde A_3+\widetilde A_4+2\widetilde A_5)(\alpha_1),\nonumber\\
F_1(\alpha_2)&=&\int_0^{1-\alpha_2}d\alpha_1(V_1-A_1)(\alpha_1,\alpha_2,1-\alpha_1-\alpha_2),\nonumber\\
F_2(\alpha_2)&=&(-\widetilde V_1+\widetilde V_2+\widetilde V_4-\widetilde A_1-\widetilde A_2-2\widetilde A_3+\widetilde A_4+2\widetilde A_5)(\alpha_2),\nonumber\\
G_1(\alpha_3)&=&\int_0^{1-\alpha_3}d\alpha_1A_1(\alpha_1,1-\alpha_1-\alpha_3,\alpha_3),\nonumber\\
G_2(\alpha_3)&=&(-\widetilde A_1+\widetilde A_2-\widetilde A_4)(\alpha_3),\nonumber\\
G_3(\alpha_3)&=&(-\widetilde A_1+\widetilde A_2-\widetilde A_3)(\alpha_3),\nonumber\\
G_4(\alpha_3)&=&(\widetilde{\widetilde A}_1-\widetilde{\widetilde
A}_2+\widetilde{\widetilde A}_3+\widetilde{\widetilde
A}_4-\widetilde{\widetilde A}_5+\widetilde{\widetilde
A}_6)(\alpha_3),
\end{eqnarray}
where
\begin{eqnarray}
\widetilde F_i(\alpha_1)&=&\int_0^{\alpha_1}d\alpha_1'\int_0^{1-\alpha_1'}d\alpha_2F_i(\alpha_1',\alpha_2,1-\alpha_1'-\alpha_2),\nonumber\\
\widetilde{\widetilde
F}_i(\alpha_1)&=&\int_0^{\alpha_1}d\alpha_1'\int_0^{\alpha_1'}d\alpha_1''\int_0^{1-\alpha_1''}d\alpha_2F_i(\alpha_1'',\alpha_2,1-\alpha_1''-\alpha_2),\nonumber\\
\widetilde
F_i(\alpha_2)&=&\int_0^{\alpha_2}d\alpha_2'\int_0^{1-\alpha_2'}d\alpha_1F_i(\alpha_1,\alpha_2',1-\alpha_1-\alpha_2'),\nonumber\\
\widetilde{\widetilde
F}_i(\alpha_2)&=&\int_0^{\alpha_2}d\alpha_2'\int_0^{\alpha_2'}d\alpha_2''\int_0^{1-\alpha_2''}d\alpha_1F_i(\alpha_1,\alpha_2'',1-\alpha_1-\alpha_2''),\nonumber\\
\widetilde F_i(\alpha_3)&=&\int_0^{\alpha_3}d\alpha_3'\int_0^{1-\alpha_3'}d\alpha_1F_i(\alpha_1,1-\alpha_1-\alpha_3',\alpha_3'),\nonumber\\
\widetilde{\widetilde
F}_i(\alpha_3)&=&\int_0^{\alpha_3}d\alpha_3'\int_0^{\alpha_3'}d\alpha_3''\int_0^{1-\alpha_3''}d\alpha_1F_i(\alpha_1,1-\alpha_1-\alpha_3'',\alpha_3'').
\end{eqnarray}
These terms come from the partial integration on $\alpha_1$, $\alpha_2$ and $\alpha_3$, which is employed to eliminate the factors $1/(P\cdot x)^n$ in the
calculation.

\subsection{LCSRs for the $\Sigma$ EM form factors}

This subsection is reserved for LCSRs of the $\Sigma^+$ EM form factors. The process for $\Sigma^-$ is similar and we only list the numerical analysis. The
calculation begins with the following correlation function:
\begin{equation} T_\mu(P,q)=i \int d^4xe^{iq\cdot x}\langle
0|T\{j_\Sigma(0)j_\mu^{em}(x)\}|\Sigma(P)\rangle,\label{correlator2}
\end{equation}
where the electromagnetic current is
\begin{equation}
j_\mu^{em}(x)=e_u\bar u(x)\gamma_\mu u(x)+e_s \bar s(x) \gamma_\mu s(x),
\end{equation}
and the interpolating current is
\begin{equation}
j_\Sigma(x)=\epsilon^{ijk}[u^i(x)C \gamma_\mu u^j(x)]\gamma_5\gamma^\mu s^k(x),
\end{equation}
with the coupling constant $ \langle 0|j_\Sigma|\Sigma\rangle=\lambda_1M\Sigma(P)$.

Following the same procedure as that in the above subsection, the sum rules are given as
\begin{eqnarray}
\lambda_1F_1(Q^2)&=&e_u\Big\{\int_{\alpha_{20}}^1d\alpha_2e^{-\frac{s-M^2}{M_B^2}}\{B_0(\alpha_2)+\frac{1}{\alpha_2}B_1(\alpha_2)
+\frac{Q^2}{\alpha_2^2M_B^2}B_2(\alpha_2)\nonumber\\
&&-\frac{M^2}{M_B^2}B_3(\alpha_2)-\frac{2M^2}{\alpha_2M_B^2}B_4(\alpha_2)\}
+e^{-\frac{s_0-M^2}{M_B^2}}\frac{\alpha_{20}^2}{\alpha_{20}^2M^2+Q^2}\nonumber\\
&&\times\{\frac{Q^2}{\alpha_{20}^2}B_2(\alpha_{20})-M^2B_3(\alpha_{20})-\frac{2M^2}{\alpha_{20}}B_4(\alpha_{20})\}\Big\}\nonumber\\
&&-e_s\Big\{\int_{\alpha_{30}}^1d\alpha_3e^{-\frac{s_1-M^2}{M_B^2}}\frac{1}{\alpha_3}\{C_0(\alpha_3)+C_1(\alpha_3)+\frac{m_s^2+\alpha_3m_sM-Q^2}{\alpha_3M_B^2}\nonumber\\
&&\times C_1(\alpha_3)-\frac{m_sM}{M_B^2}C_2(\alpha_3)-\frac{M(\alpha_3M+m_s)}{M_B^2}C_3(\alpha_3)-\frac{m_sM^3}{M_B^4}C_4(\alpha_3)\}\nonumber\\
&&-e^{-\frac{s_0-M^2}{M_B^2}}\frac{1}{\alpha_{30}^2M^2+Q^2+m_s^2}\{(m_s^2+\alpha_{30}m_sM-Q^2)C_1(\alpha_{30})\nonumber\\
&&-\alpha_{30}m_sMC_2(\alpha_{30})
-\alpha_{30}M(\alpha_{30}M+m_s)C_3(\alpha_{30})-\frac{\alpha_{30}m_sM^3}{M_B^2}\nonumber\\
&&\times C_4(\alpha_{30})+\alpha_{30}^2M^3\frac{d}{d\alpha_{30}}C_4(\alpha_{30})\frac{\alpha_{30}m_sM}{\alpha_{30}^2M^2+Q^2+m_s^2}\}\Big\},
\end{eqnarray}
and
\begin{eqnarray}
\lambda_1F_2(Q^2)&=&2e_u\Big\{\int_{\alpha_{20}}^1d\alpha_2\frac{1}{\alpha_2}\{-D_0(\alpha_2)+\frac{M^2}{M_B^2}D_1(\alpha_2)\}e^{-\frac{s-M^2}{M_B^2}}\nonumber\\
&&+e^{-\frac{s_0-M^2}{M_B^2}}\frac{\alpha_{20}M^2}{\alpha_{20}^2M^2+Q^2}D_1(\alpha_{20})\Big\}\nonumber\\
&&+2e_s\Big\{\int_{\alpha_{30}}^1\frac{1}{\alpha_3}\{E_0(\alpha_3)
+\frac{m_sM}{\alpha_3M_B^2}E_1(\alpha_3)+\frac{M^2}{M_B^2}E_2(\alpha_3)\nonumber\\
&&-\frac{m_sM^3}{\alpha_3M_B^4}E_3(\alpha_3)\}e^{-\frac{s_1-M^2}{M_B^2}}+e^{-\frac{s_0-M^2}{M_B^2}}\frac{1}{\alpha_{30}^2M^2+Q^2+m_s^2}\nonumber\\
&&\{m_sME_1(\alpha_{30})+\alpha_{30}M^2E_2(\alpha_{30})-\frac{m_sM^3}{M_B^2}E_3(\alpha_{30})\nonumber\\
&&+\alpha_{30}^2\frac{d}{d\alpha_{30}}E_(\alpha_{30}) \frac{m_sM^3}{\alpha_{30}^2M^2+Q^2+m_s^2}\}\Big\},
\end{eqnarray}
where $s=(1-\alpha_2)M^2+\frac{1-\alpha_2}{\alpha_2}Q^2$, $s_1=(1-\alpha_3)M^2+\frac{1-\alpha_3}{\alpha_3}Q^2+\frac{m_s^2}{\alpha_3}$, and the following notations
are used for convenience:
\begin{eqnarray}
B_0(\alpha_2)&=&\int_0^{1-\alpha_2}d\alpha_1(2V_1-3V_3-A_3)(\alpha_1,\alpha_2,1-\alpha_1-\alpha_2),\nonumber\\
B_1(\alpha_2)&=&(\widetilde V_1-\widetilde V_2-\widetilde V_3+\widetilde A_1-\widetilde A_2+\widetilde A_3)(\alpha_2),\nonumber\\
B_2(\alpha_2)&=&(\widetilde V_1-\widetilde V_2-\widetilde V_3-\widetilde A_1+\widetilde A_2-\widetilde A_3)(\alpha_2),\nonumber\\
B_3(\alpha_2)&=&(-2\widetilde V_1+3\widetilde V_3-\widetilde V_4+2\widetilde V_5+\widetilde A_3-\widetilde A_4)(\alpha_2),\nonumber\\
B_4(\alpha_2)&=&(-\widetilde{\widetilde V}_1+\widetilde{\widetilde V}_2+\widetilde{\widetilde V}_3+\widetilde{\widetilde V}_4+\widetilde{\widetilde V}_5-\widetilde{\widetilde V}_6)(\alpha_2),\nonumber\\
C_0(\alpha_3)&=&\int_0^{1-\alpha_3}d\alpha_1(\frac{m_s}{M}V_1+\alpha_3V_3)(\alpha_1,1-\alpha_1-\alpha_3,\alpha_3),\nonumber\\
C_1(\alpha_3)&=&(\widetilde V_1-\widetilde V_2-\widetilde V_3)(\alpha_3),\nonumber\\
C_2(\alpha_3)&=&(-\widetilde V_1+\widetilde V_3+\widetilde V_5)(\alpha_3),\nonumber\\
C_3(\alpha_3)&=&(\widetilde V_4-\widetilde V_3)(\alpha_3),\nonumber\\
C_4(\alpha_3)&=&(-\widetilde{\widetilde V}_1+\widetilde{\widetilde
V}_2+\widetilde{\widetilde V}_3+\widetilde{\widetilde
V}_4+\widetilde{\widetilde V}_5-\widetilde{\widetilde
V}_6)(\alpha_3),\nonumber\\
D_0(\alpha_2)&=&\int_0^{1-\alpha_2}d\alpha_1(V_1+A_1)(\alpha_1,\alpha_2,1-\alpha_1-\alpha_2),\nonumber\\
D_1(\alpha_2)&=&(-\widetilde V_1-\widetilde V_2+2\widetilde
V_3-\widetilde V_4+2\widetilde V_5-\widetilde A_1+\widetilde
A_2-\widetilde A_4)(\alpha_2),\nonumber\\
E_0(\alpha_3)&=&\int_0^{1-\alpha_3}d\alpha_1V_1(\alpha_1,1-\alpha_1-\alpha_3,\alpha_3),\nonumber\\
E_1(\alpha_3)&=&(\widetilde V_1-\widetilde V_2-\widetilde
V_3)(\alpha_3),\nonumber\\
E_2(\alpha_3)&=&(\widetilde V_1-\widetilde V_2-\widetilde
V_3)(\alpha_3),\nonumber\\
E_3(\alpha_3)&=&(-\widetilde{\widetilde V}_1+\widetilde{\widetilde V}_2+\widetilde{\widetilde V}_3+\widetilde{\widetilde V}_4+\widetilde{\widetilde
V}_5-\widetilde{\widetilde V}_6)(\alpha_3),
\end{eqnarray}
in which the functions with a tilde are defined the same as those in the above subsection.
\section{Numerical analysis} \label{sec3}
\subsection{Choice of the nonperturbtive parameters}
Determination of the nonperturbative parameters of the distribution amplitudes can be carried out in QCD sum rules. The parameters have been estimated in the
previous work \cite{DAs}. Here we only list the central values of the parameters.

For $\Lambda$ the sum rules give the central values of the two parameters as
\begin{eqnarray}
f_\Lambda=6.0\times10^{-3}\;\mbox{GeV}^2\;\ ,\lambda_1=1.0\times10^{-2}\; \mbox{GeV}^2.
\end{eqnarray}
For $\Sigma$ the results are:
\begin{eqnarray}
f_{\Sigma}=9.4\times10^{-3}\;\mbox{GeV}^2\;\ ,\lambda_1=-2.5\times10^{-2}\; \mbox{GeV}^2.
\end{eqnarray}
\subsection{Numerical analysis on the LCSRs}
\begin{figure}
\begin{minipage}{7cm}
\epsfxsize=5cm \centerline{\epsffile{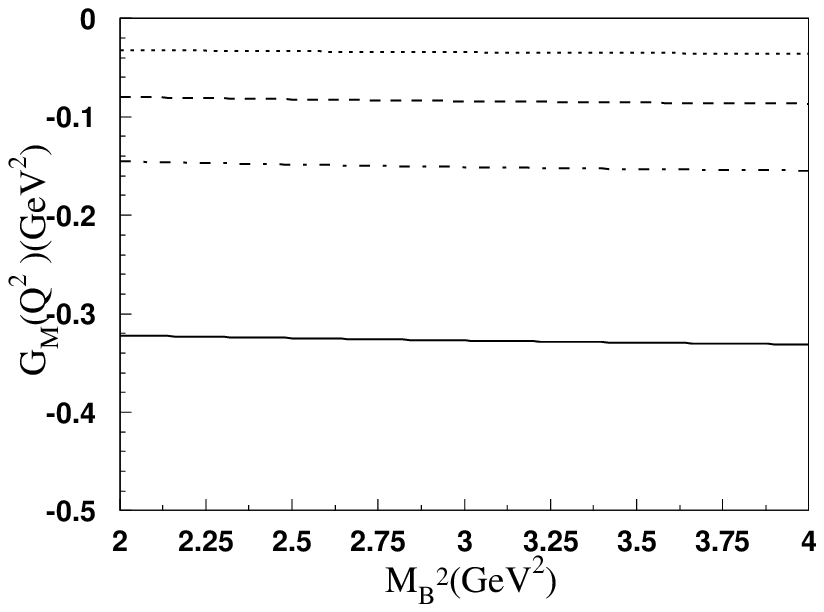}}
\end{minipage}
\hfill
\begin{minipage}{7cm}
\epsfxsize=5cm \centerline{\epsffile{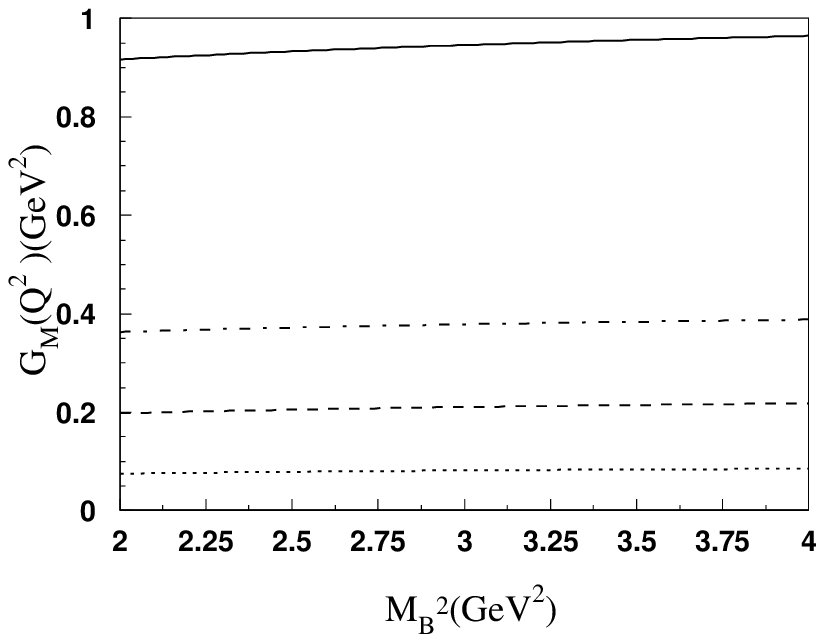}}
\end{minipage}
\caption{\it{$M_B^2$-dependence of the magnetic form factors $G_M(Q^2)$. The lines correspond to the momentum transfer at the points $Q^2=1,2,3,5\;\ \mbox{GeV}^{2}$
from the bottom up for $\Sigma^-$ (left) and from the up down for $\Sigma^+$ (right) with the threshold $s_0=2.65\;\ \mbox{GeV}^2$.}}\label{fig1}
\end{figure}

\begin{figure}
\begin{minipage}{7cm}
\epsfxsize=5cm \centerline{\epsffile{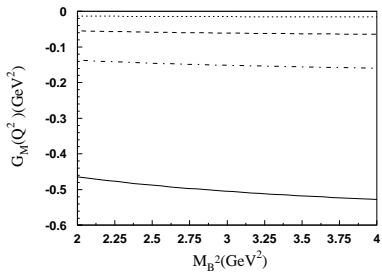}}
\end{minipage}
\caption{\it{$M_B^2$-dependence of the magnetic form factor $G_M(Q^2)$ of $\Lambda$. The lines correspond to the momentum transfer at the points $Q^2=1,2,3,5\;\
\mbox{GeV}^{2}$ from the bottom up with the threshold $s_0=2.55\;\ \mbox{GeV}^2$.}}\label{fig2}
\end{figure}

The choices of other input parameters appeared in the sum rules are as follows: the threshold is set to be $s_0=2.65-2.85\;\mbox{GeV}^2$ for $\Sigma$ and
$s_0=2.45-2.65\;\mbox{GeV}^2$ for $\Lambda$, while the mass of the hadrons are the central values provided by the particle data group (PDG) \cite{PDG}:
$M_\Lambda=1.116\;\mbox{GeV}$, $M_{\Sigma^+}=1.189\;\mbox{GeV}$ and $M_{\Sigma^-}=1.197\;\mbox{GeV}$. It is known that in the numerical analysis of the sum rules,
the auxiliary Borel parameter should have a proper region in which the physical value varies mildly with it. At the same time, the choice of the working window
needs to suppress both the higher resonance and higher twist contributions. To exhibit the working window, the magnetic form factors versus the Borel parameter at
different points of the momentum transfer $Q^2$ are plotted in Fig. \ref{fig1} (for $\Sigma$) and Fig. \ref{fig2} (for $\Lambda$), from which we can see that the
form factors are almost independent of the Borel parameter in the range $2\;\mbox{GeV}^{2}\leq M_B^2\leq 4\;\mbox{GeV}^{2}$.

\begin{figure}
\begin{minipage}{7cm}
\epsfxsize=5cm \centerline{\epsffile{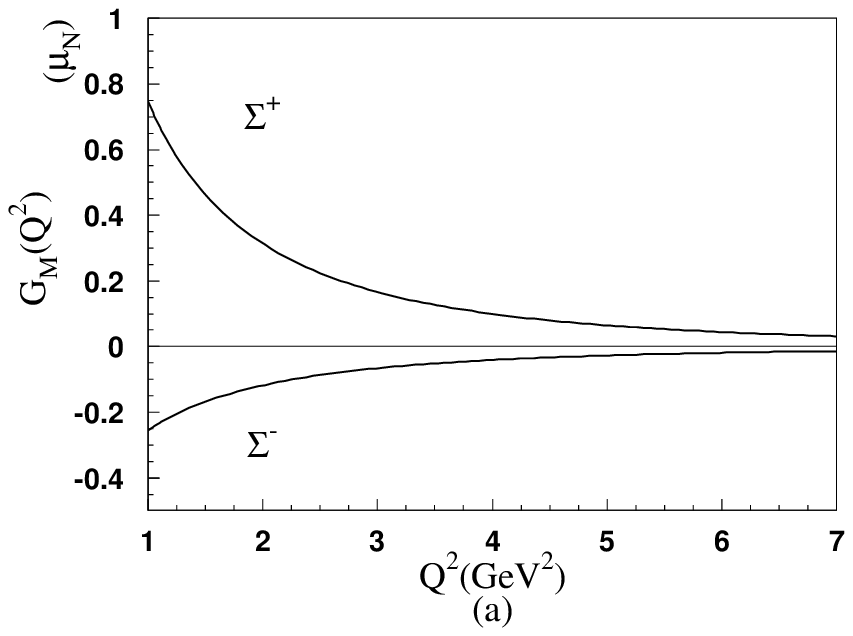}}
\end{minipage}
\hfill
\begin{minipage}{7cm}
\epsfxsize=5cm \centerline{\epsffile{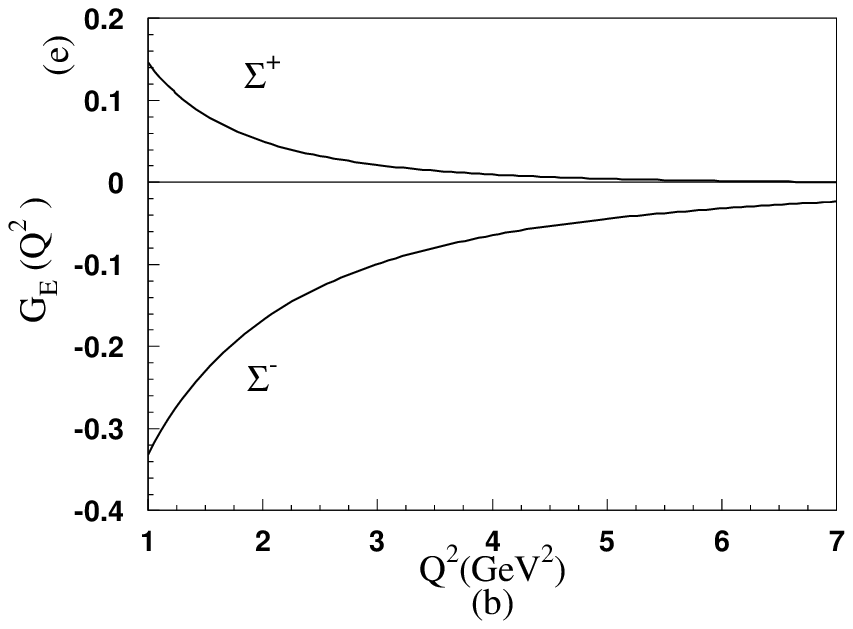}}
\end{minipage}
\caption{\it{$Q^2$-dependence of the magnetic and electric form factors of $\Sigma$. The threshold is $s_0=2.65\;\ \mbox{GeV}^2$.}}\label{fig3}
\end{figure}

Hereafter, the Borel parameter is set to be $M_B^2=3.0\;\mbox{GeV}^2$ in the numerical analysis. In Refs. \cite{Cauteren}, Van Cauteren $et$ $al.$ have given the
magnetic and electric form factors of strange baryons in the range $0\le Q^2\le6\;\mbox{GeV}^2$. Correspondingly, we plot the $Q^2$-dependent magnetic and electric
form factors of the $\Sigma$ baryons in the range $1\,\mbox{GeV}^2\le Q^2\le7\,\mbox{GeV}^2$, which are shown in Fig. \ref{fig3}. In comparison with their results,
the form factors from our approach are in accordance with theirs except the electric form factor of $\Sigma^+$, for it changes sign at some momentum transfer in
their result. As mentioned in Sec. \ref{sec2}, the normalization of the magnetic and electric form factors at point $Q^2=0$ is given by the magnetic moment and the
electric charge, respectively. However, in our frame, we can not give the behaviors of the form factors at this point directly due to the limit of LCSR method
itself. For a comparison with the results obtained from chiral quark/soliton model \cite{Kim}, in which the behaviors of the baryon electromagnetic form factors are
given in the range $0\le Q^2\le 1\,\mbox{GeV}^2$, one can see that our results agree with theirs qualitatively due to the accordance at the point
$Q^2=1\,\mbox{GeV}^2$.

\begin{figure}
\begin{minipage}{7cm}
\epsfxsize=5cm \centerline{\epsffile{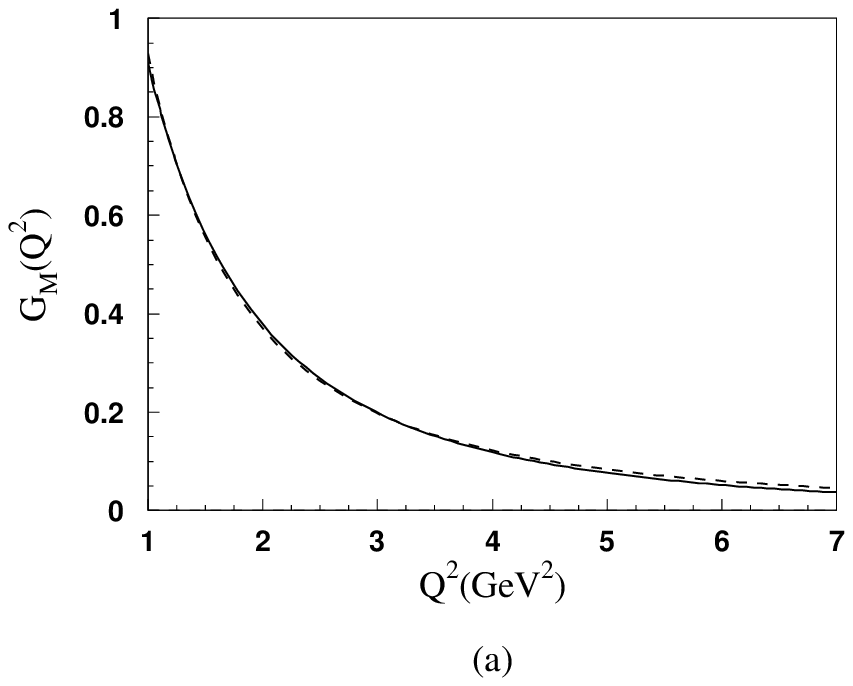}}
\end{minipage}
\hfill
\begin{minipage}{7cm}
\epsfxsize=5cm \centerline{\epsffile{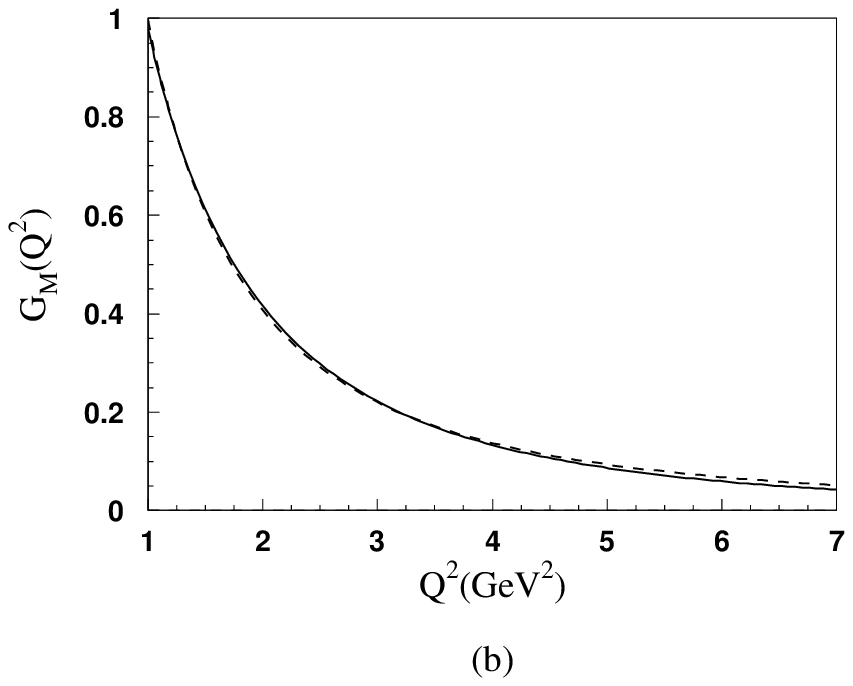}}
\end{minipage}
\caption{\it{Fits of the form factor $G_M(Q^2)$ by the dipole formula $\mu_{\Sigma^+}/(1+Q^2/m_0^2)^2$. The dashed lines are the fits, and figures $(a),(b)$
correspond to the threshold $s_0=2.65,2.85\; \mbox{GeV}^2$, respectively.}}\label{fig4}
\end{figure}

\begin{figure}
\begin{minipage}{7cm}
\epsfxsize=5cm \centerline{\epsffile{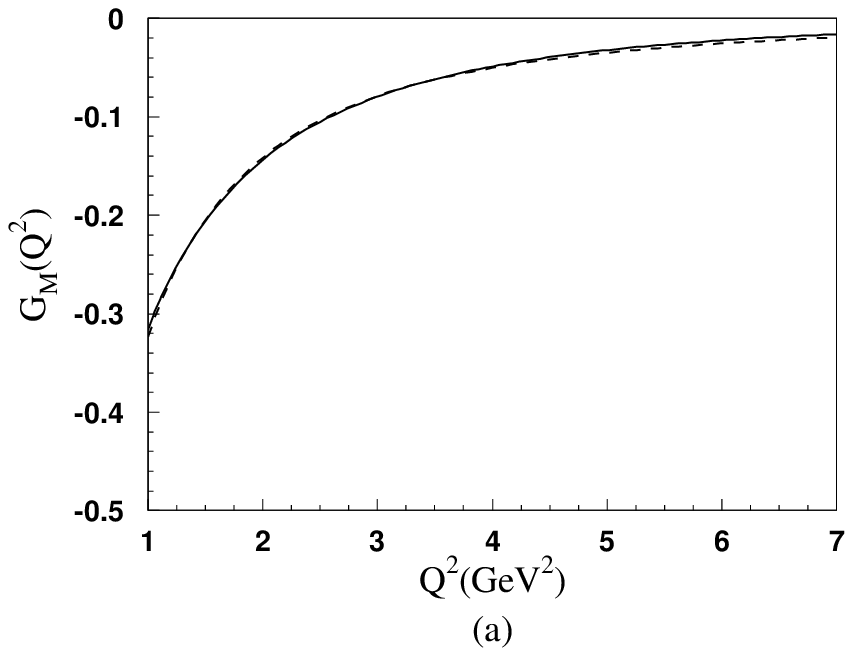}}
\end{minipage}
\hfill
\begin{minipage}{7cm}
\epsfxsize=5cm \centerline{\epsffile{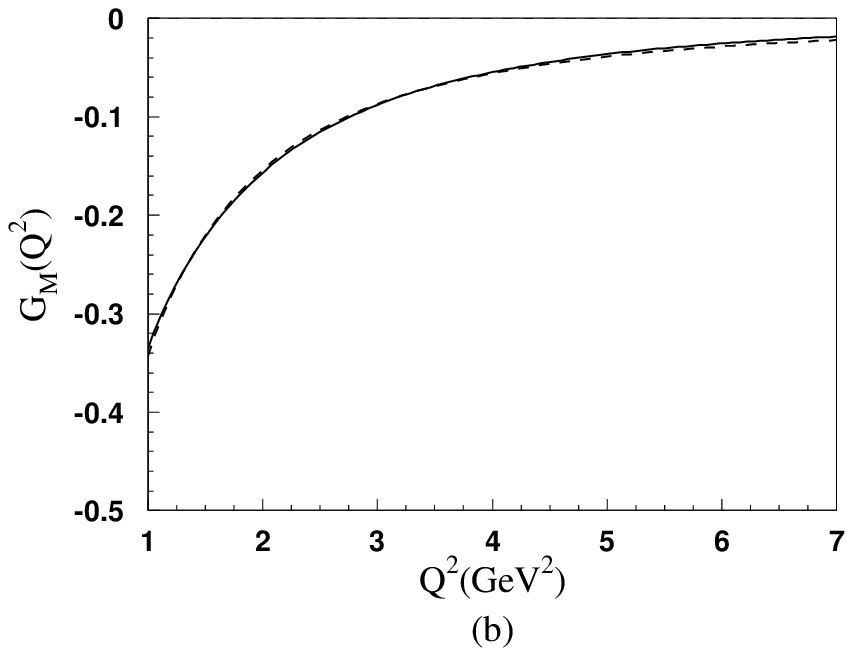}}
\end{minipage}
\caption{\it{Fits of the form factor $G_M(Q^2)$ by the dipole formula $\mu_{\Sigma^-}/(1+Q^2/m_0^2)^2$. The dashed lines are the fits, and figures $(a),(b)$
correspond to the threshold $s_0=2.65,2.85\; \mbox{GeV}^2$, respectively.}}\label{fig5}
\end{figure}

It can be seen from the figure that the magnetic form factors approach zero with the increase of the momentum transfer, which leads to the assumption that the
magnetic form factors can be described by the dipole formula:
\begin{equation}\label{dipole}
\frac{1}\mu G_M(Q^2)=\frac{1}{(1+Q^2/m_0^2)^2}=G_D(Q^2),
\end{equation}
where $\mu$ corresponds to the baryon magnetic moment. To test our calculation, we estimate the magnetic moments of the baryons from the sum rules. The similar
process is adopted as that in Ref. \cite{DAs}, in which the magnetic form factor is fitted by the formula $\mu G_D(Q^2)$ in the sum rule allowed range
$1\,\mbox{GeV}^2\le Q^2\le7\,\mbox{GeV}^2$, and the magnetic moments are estimated from the fits. Fig. \ref{fig4} (Fig. \ref{fig5}) is the dipole formula fit of the
magnetic form factor $G_M(Q^2)$ for $\Sigma^+$ ($\Sigma^-$). The numerical estimations are: $\mu_{\Sigma^+}=(4.19\pm0.04)\mu_N$ and
$\mu_{\Sigma^-}=-(1.05\pm0.01)\mu_N$.

\begin{figure}
\begin{minipage}{7cm}
\epsfxsize=5cm \centerline{\epsffile{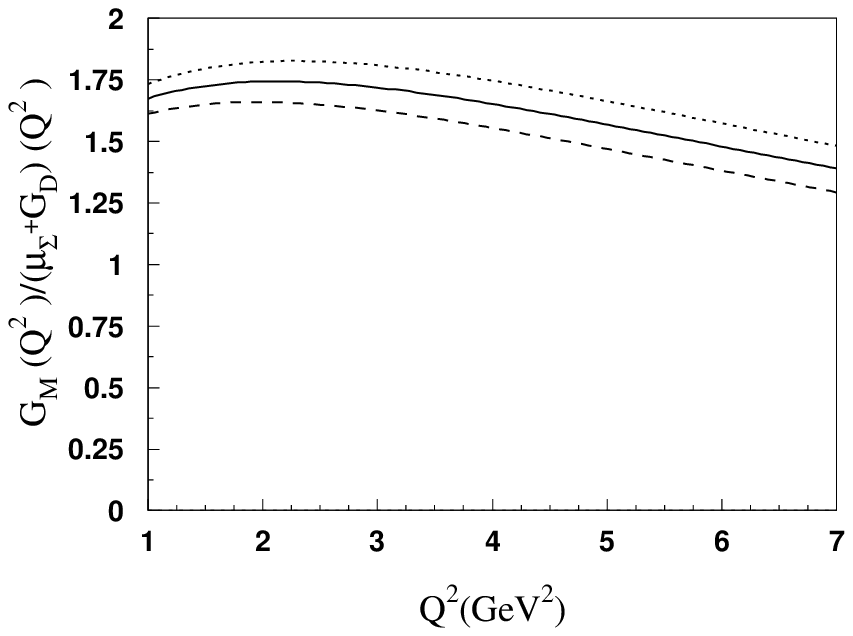}}
\end{minipage}
\hfill
\begin{minipage}{7cm}
\epsfxsize=5cm \centerline{\epsffile{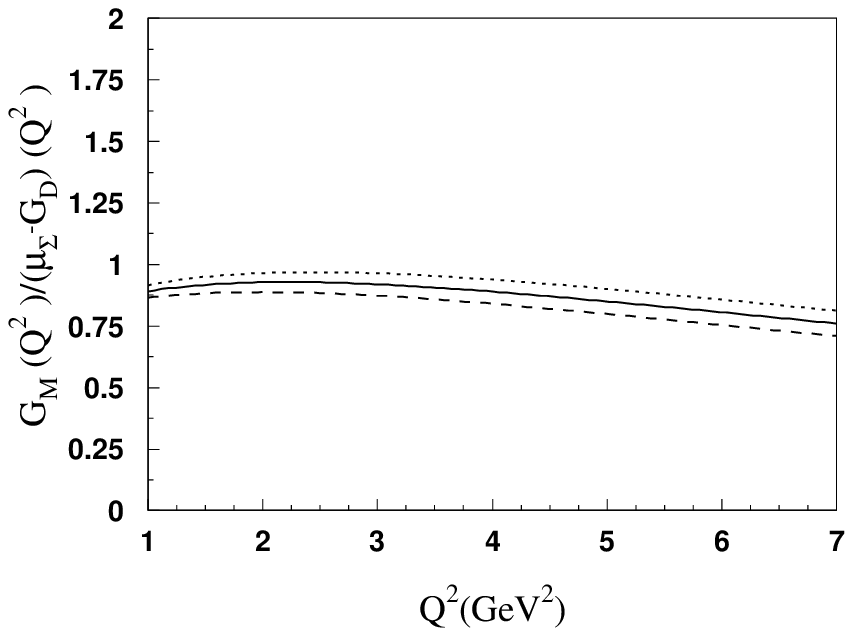}}
\end{minipage}
\caption{\it{$Q^2$-dependence of $G_M(Q^2)/(\mu_\Sigma G_D(Q^2))$. The three lines correspond to the threshold $s_0=2.65,2.75,2.85\;\mbox{GeV}^2$ from the bottom
up.}}\label{fig6}
\end{figure}

The $Q^2$-dependence of the physical value $G_M(Q^2)/(\mu_\Sigma G_D(Q^2))$ is shown in Fig. \ref{fig6}. In the analysis the magnetic moment is from the PDG
\cite{PDG}: $\mu_{\Sigma^+}=2.458\mu_N$ and $\mu_{\Sigma^-}=-1.160\mu_N$. The other parameter $m_0^2$ is the central value from the dipole formula fit that is
$m_0^2=0.74\;\mbox{GeV}^2$ for $\Sigma^+$ and $m_0^2=0.99\;\mbox{GeV}^2$ for $\Sigma^-$.

\begin{figure}
\begin{minipage}{7cm}
\epsfxsize=5cm \centerline{\epsffile{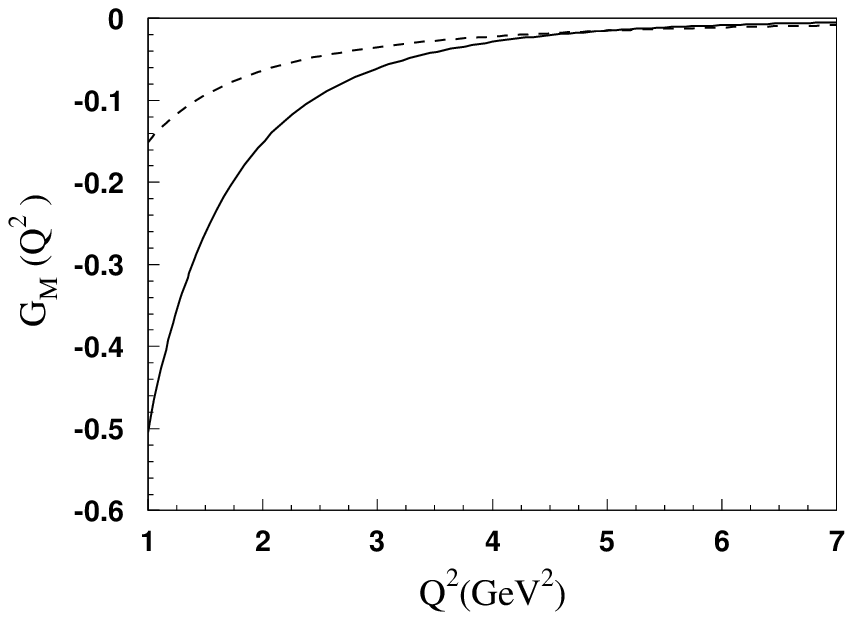}}
\end{minipage}
\hfill
\begin{minipage}{7cm}
\epsfxsize=5cm \centerline{\epsffile{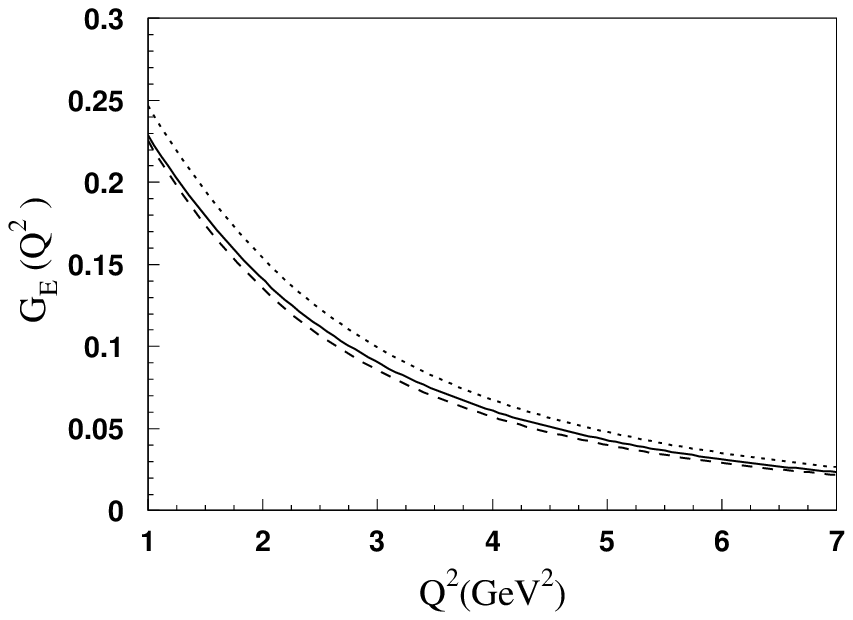}}
\end{minipage}
\caption{\it{$Q^2$-dependence of magnetic (left) and electric (right) form factors of $\Lambda$.}}\label{fig7}
\end{figure}

\begin{figure}
\begin{minipage}{7cm}
\epsfxsize=5cm \centerline{\epsffile{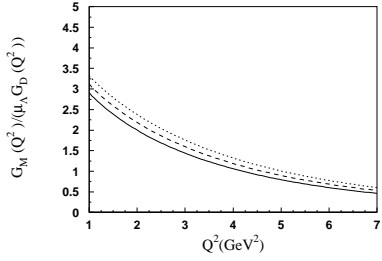}}
\end{minipage}
\caption{\it{$Q^2$-dependence of $G_M(Q^2)/(\mu_\Lambda G_D(Q^2))$ on $\Lambda$. The three lines correspond to the threshold $s_0=2.45,2.55,2.65\;\mbox{GeV}^2$ from
the bottom up.}}\label{fig8}
\end{figure}

The same process is carried out for the $\Lambda$ baryon. The analysis shows that unlike results from Ref. \cite{EMff}, in which the magnetic moment can be well
estimated from the fit, the magnetic form factor obtained with the adoption of the Ioffe type current fails to be fitted by the dipole formula assumption. Herein we
only present the $Q^2$-dependence of the magnetic and electric form factors in Fig. \ref{fig7}. The magnetic form factor is plotted in the left one, in which the
dashed line is the dipole formula fit from Ref. \cite{EMff}. The figure shows that the magnetic form factor approaches zero faster than the dipole formula with the
increase of $Q^2$, which is different from the results of the nucleon from the polarization technic experiments \cite{Jlab}. In addition, we give the physical value
$G_M(Q^2)/(\mu_\Lambda G_D(Q^2))$ on the momentum transfer $Q^2$ in Fig. \ref{fig8}, where the parameters used in the dipole formula are chosen as follows: the
magnetic moment is from PDG that is $\mu_\Lambda=-0.613\mu_N$ and the other parameter $m_0^2$ is from the previous work \cite{EMff}: $m_0^2=0.89$. One can also see
from Fig. \ref{fig8} that the dipole formula assumption fails to describe the magnetic form factor.

We list results from the two different interpolating currents in Table \ref{tab}, from which we can see that the adoption of the Ioffe type current is good for the
estimation of the magnetic moment of $\Sigma^-$ but fails to estimate that of $\Lambda$. However, the utilization of the CZ current gives a good estimation on the
$\Lambda$ magnetic moment. More accurate estimation of the magnetic moments of the baryons needs to consider the higher order conformal spin corrections to the
distribution amplitudes. Besides the effect from the choice of the interpolating current, the higher order QCD coupling $\alpha_s$ corrections may affect the result
to some extent.
\begin{table}[h]
\caption{The magnetic moment of the baryons from various interpolating currents.}
\begin{center}
\begin{tabular}{|c|c|c|c|}
\hline
$\mu$($\mu_N$)&$\Sigma^+$&$\Sigma^-$&$\Lambda$\\
\hline
CZ current& $3.13\pm0.11$ & $-1.59\pm0.02$ &$0.64\pm0.04$ \\
\hline
Ioffe current& $4.19\pm0.04$ & $-1.05\pm0.01$ &/ \\
\hline
PDG& $2.458\pm0.010$ & $-1.160\pm0.025$ & $-0.613\pm0.004$ \\
\hline
\end{tabular}
\end{center}\label{tab}
\end{table}

In summary, we investigate the EM form factors of $\Lambda$ and $\Sigma$ to leading order in QCD within the framework of LCSR by using Ioffe type interpolating
currents. The magnetic form factors are fitted by the dipole formula, from which the magnetic moments of the baryons are estimated as:
$\mu_{\Sigma^+}=(4.19\pm0.04)\mu_N$ and $\mu_{\Sigma^-}=-(1.05\pm0.01)\mu_N$. In comparison with the previous work \cite{DAs,EMff}, the dipole formula assumption
has a good estimation for $\Sigma^-$, but fails to estimate the magnetic moment of $\Lambda$. This may partly lie in the fact that the vector like structures of the
$\Lambda$ distribution amplitudes can play more important roles in the calculations with the choice of the Ioffe type current, which needs to consider higher order
conformal spin corrections. It can be concluded that the magnetic form factor of $\Lambda$ from the Ioffe type current approaches zero faster than that of the
dipole formula with the momentum transfer $Q^2$. Future experiments are expected to test the calculations and give us more information on the electromagnetic form
factors.

\acknowledgments  This work was supported in part by the National
Natural Science Foundation of China under Contract No.10675167.
\appendix
\section{Distribution amplitudes}\label{appendix}
The distribution amplitudes of the $\Lambda$ and $\Sigma$ baryons have been given up to twist $6$ in the conformal expansion to leading order conformal spin
accuracy \cite{Wang,DAs}. In this section we only list them out for the completeness of this paper. As in the calculations only vector and axial-vector like
distribution amplitudes contribute, the following structure terms are given for simplicity \cite{Braun,Wang,DAs}:
\begin{eqnarray}
&& 4 \langle{0} |\epsilon^{ijk} {q_1}_\alpha^i(a_1 x) {q_2}_\beta^j(a_2 x) s_\gamma^k(a_3 x) |{B(P)}\rangle
\nonumber \\
&=& {\cal V}_1  \left(\!\not\!{P}C \right)_{\alpha \beta} \left(\gamma_5 B\right)_\gamma + {\cal V}_2 M \left(\!\not\!{P} C \right)_{\alpha \beta} \left(\!\not\!{x}
\gamma_5 B\right)_\gamma  + {\cal V}_3 M  \left(\gamma_\mu C \right)_{\alpha \beta}\left(\gamma^{\mu} \gamma_5 B\right)_\gamma
\nonumber \\
&& + {\cal V}_4 M^2 \left(\!\not\!{x}C \right)_{\alpha \beta} \left(\gamma_5 B\right)_\gamma + {\cal V}_5 M^2 \left(\gamma_\mu C \right)_{\alpha \beta} \left(i
\sigma^{\mu\nu} x_\nu \gamma_5 B\right)_\gamma + {\cal V}_6 M^3 \left(\!\not\!{x} C \right)_{\alpha \beta} \left(\!\not\!{x} \gamma_5 B\right)_\gamma
\nonumber \\
&& + {\cal A}_1  \left(\!\not\!{P}\gamma_5 C \right)_{\alpha \beta} B_\gamma + {\cal A}_2 M \left(\!\not\!{P}\gamma_5 C \right)_{\alpha \beta} \left(\!\not\!{x}
B\right)_\gamma  + {\cal A}_3 M \left(\gamma_\mu \gamma_5 C \right)_{\alpha \beta}\left( \gamma^{\mu} B\right)_\gamma
\nonumber \\
&& + {\cal A}_4 M^2 \left(\!\not\!{x} \gamma_5 C \right)_{\alpha \beta} B_\gamma + {\cal A}_5 M^2 \left(\gamma_\mu \gamma_5 C \right)_{\alpha \beta} \left(i
\sigma^{\mu\nu} x_\nu B\right)_\gamma + {\cal A}_6 M^3 \left(\!\not\!{x} \gamma_5 C \right)_{\alpha \beta} \left(\!\not\!{x} B\right)_\gamma \,,\label{da-def}
\end{eqnarray}
where $B_\gamma$ is the spinor of the baryon, $q_i$ represent $u$ or $d$ quark fields, $a_i$ are real numbers denoting coordinates of valence quarks, $C$ is the
charge conjugation matrix and $\sigma_{\mu\nu}=\frac{i}{2}[\gamma_\mu,\gamma_\nu]$. All the functions  $\mathcal {A}_i$ and $\mathcal{V}_i$ depend on the scalar
product $P\cdot x$.

These calligraphic invariant functions do not have a definite twist, thus we need to express them in terms of the distribution amplitudes $V_i$ and $A_i$ with a
definite twist. The following relations hold for vector distributions:
 \begin{eqnarray}
\renewcommand{\arraystretch}{1.7}
\begin{array}{lll}
 \mathcal V_1 = V_1\,, &~~& 2 p\cdot x \mathcal V_2 = V_1 - V_2 - V_3\,, \\
 2 \mathcal V_3 = V_3\,, &~~& 4 p\cdot x \mathcal V_4 = - 2 V_1 + V_3 + V_4  + 2 V_5\,, \\
4 p\cdot x \mathcal V_5 = V_4 - V_3\,, &~~& (2 p\cdot x )^2 \mathcal V_6 = - V_1 + V_2 +  V_3 +  V_4 + V_5 - V_6\,,
\end{array}
\renewcommand{\arraystretch}{1.0}
\end{eqnarray}
and for axial vector distributions:
 \begin{eqnarray}
\renewcommand{\arraystretch}{1.7}
\begin{array}{lll}
 \mathcal A_1 = A_1\,, &~~& 2 p\cdot x \mathcal A_2 = - A_1 + A_2 -  A_3\,, \\
 2 \mathcal A_3 = A_3\,, &~~& 4 p\cdot x \mathcal A_4 = - 2 A_1 - A_3 - A_4  + 2 A_5\,, \\
 4 p\cdot x \mathcal A_5 = A_3 - A_4\,, &~~&
(2 p\cdot x )^2  \mathcal A_6 =  A_1 - A_2 +  A_3 +  A_4 - A_5 + A_6\,.
\end{array}
\renewcommand{\arraystretch}{1.0}
\end{eqnarray}
To leading order conformal spin accuracy, the above distribution amplitudes can be expressed by the coupling constants $f_B$ and $\lambda_1$, which are defined as
\begin{eqnarray}
\langle 0|j_{CZ}(0)|B(P)\rangle &=& f_B(P\cdot z)\not\!zB(P),\nonumber \\
\langle 0|j_B(0)|B(P)\rangle &=& \lambda_1MB(P),
\end{eqnarray}
where $j_B$ is the Ioffe type current of the baryon $B$.

For $\Lambda$, the distribution amplitudes can be expanded to leading order conformal spin accuracy:
\begin{eqnarray}
V_1(x_i)&=&0,\hspace{4.5cm}A_1(x_i)=-120x_1x_2x_3\phi_3^0
\end{eqnarray}
for twist-$3$ and
\begin{eqnarray}
V_2(x_i)&=&0,\hspace{5.0cm}A_2(x_i)=-24x_1x_2\phi_4^0,\nonumber\\
V_3(x_i)&=&12(x_1-x_2)x_3\psi_4^0,\hspace{2.3cm}A_3(x_i)=-12x_3(1-x_3)\psi_4^0
\end{eqnarray}
for twist-4 and
\begin{eqnarray}
V_4(x_i)&=&3(x_2-x_1)\psi_5^0,\hspace{2.8cm}A_4(x_i)=-3(1-x_3)\psi_5^0,\nonumber\\
V_5(x_i)&=&0,\hspace{4.9cm}A_5(x_i)=-6x_3\phi_5^0
\end{eqnarray}
for twist-5 and
\begin{eqnarray}
V_6(x_i)&=&0,\hspace{4.5cm}A_6(x_i)=-2\phi_6^0
\end{eqnarray}
for twist-6 distribution amplitudes. The six nonperturbative parameters can be expressed by $f_\Lambda$ and $\lambda_1$:
\begin{eqnarray}
\phi_3^0&=&\phi_6^0=-f_\Lambda,\hspace{2.8cm}\phi_4^0=\phi_5^0=-\frac12(f_\Lambda+\lambda_1),\nonumber\\
\psi_4^0&=&\psi_5^0=\frac12(f_\Lambda-\lambda_1).
\end{eqnarray}

Similar for the $\Sigma$ baryon distribution amplitudes:
\begin{eqnarray}
V_1(x_i)&=&120x_1x_2x_3\phi_3^0,\hspace{2.5cm}A_1(x_i)=0
\end{eqnarray}
for twist-$3$ and
\begin{eqnarray}
V_2(x_i)&=&24x_1x_2\phi_4^0,\hspace{3.9cm}A_2(x_i)=0,\nonumber\\
V_3(x_i)&=&12x_3(1-x_3)\psi_4^0,\hspace{2.8cm}A_3(x_i)=-12x_3(x_1-x_2)\psi_4^0
\end{eqnarray}
for twist-$4$ and
\begin{eqnarray}
V_4(x_i)&=&3(1-x_3)\psi_5^0,\hspace{2.95cm}A_4(x_i)=3(x_1-x_2)\psi_5^0,\nonumber\\
V_5(x_i)&=&6x_3\phi_5^0,\hspace{4.05cm}A_5(x_i)=0
\end{eqnarray}
for twist-$5$ and finally
\begin{eqnarray}
V_6(x_i)&=&2\phi_6^0,\hspace{2.5cm}A_6(x_i)=0
\end{eqnarray}
for twist-$6$ distribution amplitudes. The nonperturbative parameters are expressed as follows:
\begin{eqnarray}
\phi_3^0&=&\phi_6^0=f_{\Sigma},\hspace{2.8cm}\psi_4^0=\psi_5^0=\frac12(f_{\Sigma}-\lambda_1),\nonumber\\
\phi_4^0&=&\phi_5^0=\frac12(f_{\Sigma}+\lambda_1).
\end{eqnarray}

\end{document}